%% file: AA44_Gaensler.tex
\documentclass{ar2e_mod}

\usepackage{ar-patch,multicol,ARAstroBib,natbib}

\input psfig.sty
\jname{Annu. Rev. Astron. Astrophys.}
\jyear{2006}
\jvol{Volume 44}

\renewcommand{\textheight}{48pc}
\renewcommand{\topmargin}{7pc}

\newcommand\la{\mathrel{\hbox{\rlap{\hbox{\lower4pt\hbox{$\sim$}}}\hbox{$<$}}}}
\newcommand\ga{\mathrel{\hbox{\rlap{\hbox{\lower4pt\hbox{$\sim$}}}\hbox{$>$}}}}

\def\kms{km~s$^{-1}$}
\def\etal{{\rm et~al.\ }}

\def\farcm{\hbox{$.\mkern-4mu^\prime$}}

\begin{document}
\ifx\href\undefined\else\hypersetup{linktocpage=true}\fi 

\markboth{\sc Gaensler \& Slane}{\sc Pulsar Wind Nebulae}
\title{\sc {\bf{\sc The Evolution and Structure of \\ Pulsar Wind Nebulae \\
 \rule{\textwidth}{1mm}}}}

\author{{\Large Bryan M.\ Gaensler and Patrick O.\ Slane}
\affiliation{Harvard-Smithsonian Center for
Astrophysics, \\ 60 Garden Street, 
Cambridge, Massachusetts 02138, USA; \\
email: bgaensler@cfa.harvard.edu, pslane@cfa.harvard.edu}}



\begin{abstract}
\input abstract.tex
\end{abstract}

\maketitle

\section{INTRODUCTION}
\label{sec_intro}

\input sec1.tex

\section{OVERALL PROPERTIES}
\label{sec_overall}

\input sec2.tex

\section{PULSAR WIND NEBULA EVOLUTION}
\label{sec_evol}

\input sec3.tex

\section{YOUNG PULSAR WIND NEBULAE}
\label{sec_young}

\input sec4.tex

\section{BOW SHOCKS AROUND HIGH VELOCITY PULSARS}
\label{sec_bow}

\input sec5.tex

\section{OTHER TOPICS AND RECENT RESULTS}
\label{sec_recent}

\input sec6.tex

\input summary.tex

\vspace{.4in}
\noindent
{\bf ACKNOWLEDGMENTS}
\vspace{.1in}

\noindent We thank NASA for generous support through the
LTSA and General Observer programs, and the Radcliffe Institute 
for hosting a stimulating PWN workshop in 2005.
We also acknowledge support from an Alfred P.\ Sloan
Research Fellowship (BMG) and from
NASA Contract NAS8-03060 (POS).
We have used images provided by Tracey DeLaney, Bruno
Kh\'elifi, CXC/SAO, NRAO/AUI and ESO.
We thank Shami Chatterjee, Roger Chevalier,
Maxim Lyutikov and Steve Reynolds 
for useful discussions and suggestions.  Finally, we
thank our other enthusiastic collaborators on PWNe, most notably Jon
Arons, Rino Bandiera, Fernando Camilo, Yosi Gelfand, David Helfand,
Jack Hughes, Vicky Kaspi, Fred Seward,
Ben Stappers and Eric van der Swaluw.

\vspace{.5in}

\input AA44_Gaensler.bbl
\hrule width \columnwidth

\section*{RELATED REVIEWS}

\noindent
{\bf Supernovae and Supernova Remnants} \\
K W Weiler, R A Sramek \\
Annual Review of Astronomy and Astrophysics, Volume 26,
Page 295--341, February 1988 

\smallskip
\noindent
{\bf Recent Developments Concerning the Crab Nebula} \\
K Davidson, R A Fesen \\
Annual Review of Astronomy and Astrophysics, Volume 23,
Page 119--146, February 1985 

\smallskip
\noindent
{\bf Supernova Remnants} \\
L Woltjer \\
Annual Review of Astronomy and Astrophysics, Volume 10,
Page 129--158, February 1972 

\hrule width \columnwidth

\section*{LINKS}

\noindent
http://www.mrao.cam.ac.uk/surveys/snrs/ \\
http://www.atnf.csiro.au/research/pulsar/psrcat/ \\
http://www.physics.mcgill.ca/$\sim$pulsar/pwncat.html \\

\end{document}

%% file: abstract.tex
Pulsars steadily dissipate their rotational energy via relativistic
winds. Confinement of these outflows generates luminous pulsar wind
nebulae, seen across the electromagnetic spectrum in synchrotron and
inverse Compton emission, and in optical emission lines when they 
shock the surrounding medium.  These sources act as important
probes of relativistic shocks, particle acceleration and of interstellar
gas.  We review the many recent advances in the study of pulsar wind
nebulae, with particular focus on the evolutionary stages through which
these objects progress as they expand into their surroundings, and on
morphological structures within these nebulae which directly trace the
physical processes of particle acceleration and outflow. We conclude by
considering some exciting new probes of pulsar wind nebulae, including
the study of TeV gamma-ray emission from these sources, and observations
of pulsar winds in close binary systems.

%% file: sec1.tex
\begin{figure}[t!]
\centerline{\psfig{file=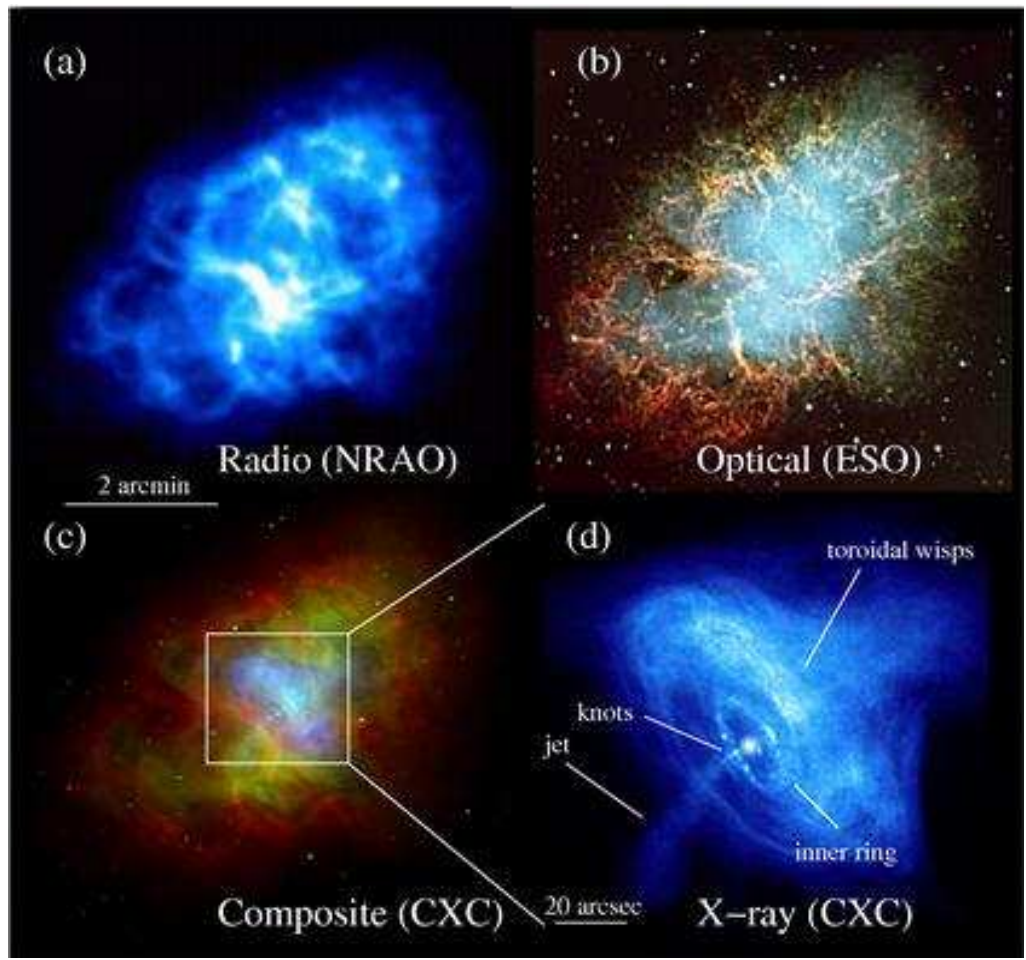,width=\textwidth}}
\caption{Images of the Crab Nebula (G184.6--5.8). 
(a)  Radio synchrotron emission
from the confined wind, with enhancements along filaments.  (b)
Optical synchrotron emission (blue-green) surrounded by emission
lines from filaments (red).  (c) Composite image of radio (red),
optical (green) and X-ray emission (blue).  (d) X-ray synchrotron
emission from jets and wind downstream of the termination shock,
marked by the inner ring.  Note the decreasing size of the synchrotron
nebula going from the radio to the X-ray band.  Each image is
oriented with north up and east to the left. The scale is indicated
by the 2 arcmin scale bar, except for panel (d), where the 20 arcsec
scale bar applies.}
\label{fig_crab}
\end{figure}

\marginnote{{\bf Neutron star:} a compact
degenerate stellar remnant, formed in the core-collapse
of a massive star. \\
{\bf Pulsar:} A rapidly rotating, highly-magnetized
neutron star, which generates coherent beams of
radiation along its magnetic poles. \\
{\bf Pulsar wind nebula:} A bubble of shocked
relativistic particles, produced when a pulsar's
relativistic wind interacts with its environment.}
The Crab Nebula (Fig.~\ref{fig_crab}) is almost certainly associated 
with a supernova (SN) explosion observed in 1054 CE
\citep[][and references therein]{sg02b}. However, this source differs
substantially from what is now seen at the sites of other recent SNe,
in that the Crab Nebula is centrally filled at all wavelengths, while
sources such as Tycho's and Kepler's supernova remnants (SNRs) show a
shell morphology.  This and other simple observations show that the Crab
Nebula is anomalous, its energetics dominated by continuous injection
of magnetic fields and relativistic particles from a central source.

A 16th magnitude star embedded in the Crab Nebula was long presumed to
be the stellar remnant and  central engine \citep{min42,pac67}.  This was
confirmed when 33-ms optical and radio pulsations were detected from this
star in the late 1960s \citep{sr68,cdt69}, and these pulsations were then
shown to be slowing down at a rate of 36~ns per day \citep{rc69b}.
The conclusion was quickly reached that the Crab Nebula contains a
rapidly rotating young neutron star, or ``pulsar'', formed in the SN of
1054 CE.  The observed rate of spin down implies that kinetic energy
is being dissipated at a rate of $\sim5\times10^{38}$~ergs~s$^{-1}$,
a value similar to the inferred rate at which energy is being supplied
to the nebula \citep{gol69}.  Following this discovery,  a theoretical
understanding was soon developed in which the central pulsar generates
a magnetized particle wind, whose ultrarelativistic electrons and
positrons radiate synchrotron emission across the electromagnetic
spectrum \citep{ps73,rg74}.  The pulsar has steadily released about a
third of its total reservoir of $\sim5\times10^{49}$~ergs of rotational
energy into its surrounding nebula over the last 950 years.
This is in sharp contrast to shell-like SNRs, in which the dominant
energy source is the $\sim10^{51}$~ergs of kinetic energy released at
the moment of the original SN explosion.

Observations over the last several decades have identified 40 to
50 further sources, in both our own Galaxy and in the Magellanic
Clouds, with properties similar to those of the Crab Nebula
\citep{gre04,krh04} --- these sources are known as ``pulsar wind
nebulae'' (PWNe).\footnote{PWNe are also often referred to as
``plerions''. However, given this term's obscure origin \citep{wp78,sha79},
we avoid using this terminology here.} Sometimes a PWN is surrounded
by a shell-like SNR, and the system is termed ``composite''
(see Fig.~\ref{fig_g21}).
In other cases, best typified by the Crab itself,
no surrounding shell is seen.\footnote{The absence
of a shell around the
Crab Nebula is presumably because it has
not yet interacted with sufficient surrounding gas \citep{fkcg95,sgs05}.}

\marginnote{{\bf SN(e)}: Supernova(e) \\ 
{\bf SNR:} Supernova remnant \\ 
{\bf PWN(e):} Pulsar wind nebula(e) \\ 
{\bf ISM:} Interstellar medium \\
{\bf MHD:} Magnetohydrodynamic \\
{\bf IC:} Inverse Compton} 
More recently, an additional category of PWNe has been identified,
in which pulsars with high space velocities produce nebulae with
cometary or bow shock morphologies as they move through the
interstellar medium (ISM) at supersonic speeds. 
The sample of such sources is currently small, but high spatial
resolution observations, especially in the X-ray band, are rapidly
adding to this group.

Because PWNe have a well-defined central energy source and are close
enough to be spatially resolved, they act as a marvelous testing ground
for studying both relativistic flows and the shocks that result when these
winds collide with their surroundings.  
Studies of PWNe, particularly the spectacular
images now being taken by the {\em Chandra X-ray Observatory}, allow us
to resolve details of the interaction of relativistic flows with their
surroundings that may never be possible in other classes of source,
and can provide the physical foundation for understanding a wide range
of astrophysical problems.

We here review current understanding of the structure and
evolution of pulsar wind nebulae, with an emphasis on the explosion of
new data and new ideas that have emerged in the last few years. Our focus
is primarily observational; theoretical considerations
have been recently discussed by
\cite{vdk04}, \cite{mel04} and \cite{che05}.

The outline of this review is as follows: in \S\ref{sec_overall} we
explain the basic observational properties of pulsars and their nebulae;
in \S\ref{sec_evol} we review current understanding of the evolutionary
sequence spanned by the observed population of PWNe;  in \S\ref{sec_young}
we discuss observations of PWNe around young pulsars, which represent the
most luminous and most intensively studied component of the population;
in \S\ref{sec_bow} we consider the properties of the bow shocks produced
by high velocity pulsars; and in \S\ref{sec_recent} we briefly
describe other recent and interesting results in this field.

%% file: sec2.tex
\subsection{Pulsar Spin Down}

Since a pulsar's rotational energy, $E_{rot}$, is the source for
most of the emission seen from PWNe, we first consider the spin evolution
of young neutron stars.

\subsubsection{Spin-Down Luminosity, Age and Magnetic Field}

An isolated pulsar has a spin period, $P$, and a period derivative with
respect to time, $\dot{P} \equiv dP/dt$, both of which can be determined
from observations of the pulsed signal.

The ``spin down luminosity'' of the pulsar, $\dot{E} = -dE_{rot}/dt$,
is the rate at which rotational kinetic energy is dissipated, and
is thus given by the equation:
\begin{equation}
\dot{E} \equiv 4 \pi^2 I \frac{\dot{P}}{P^3},
\end{equation}
where $I$ is the neutron star's moment of inertia and is usually
assumed to have the value $10^{45}$~g~cm$^{-2}$. Values of $\dot{E}$
for the observed pulsar population range between
$\approx5\times10^{38}$~ergs~s$^{-1}$ for the Crab pulsar and PSR~J0537--6910,
down to $3\times10^{28}$~ergs~s$^{-1}$
for the slowest known pulsar, PSR~J2144--3933 \citep{mhth05}.
Typically only pulsars with $\dot{E} \ga 4\times10^{36}$~ergs~s$^{-1}$ 
(of which $\sim15$ are currently known) produce prominent PWNe \citep{got04}.

The age and surface magnetic field strength of a neutron star can
be inferred from $P$ and $\dot{P}$, subject to certain assumptions.
If a pulsar spins down from an initial spin period
$P_0$ such that $\dot{\Omega} = - k \Omega^n$ (where
$\Omega = 2\pi/P$ and $n$ is the ``braking index''), 
then the age of the system is \citep{mt77}:
\begin{equation}
\tau = \frac{P}{(n-1)\dot{P}} \left[ 1 - \left(\frac{P_0}{P}\right)^{n-1} \right],
\label{eqn_tau}
\end{equation}
where we have assumed
$k$ to be a constant and $n\neq1$.
The braking index, $n$, has only been confidently measured
for four pulsars \citep[][and references therein]{lkg05},
in each case falling in the range $2< n < 3$.

If for the rest of the population we assume
$n=3$ (corresponding to spin down via magnetic
dipole radiation) and $P_0 \ll P$, Equation~(\ref{eqn_tau})
reduces to the expression for the ``characteristic age'' of a pulsar,
\begin{equation}
\tau_c \equiv \frac{P}{2\dot{P}}.
\label{eqn_tauc}
\end{equation}
Equation~(\ref{eqn_tauc}) often overestimates the true age
of the system, indicating that $P_0$ is not much smaller than $P$
\citep[e.g.,][]{mgb+02}.
PWNe resembling the Crab Nebula tend to be observed only
for pulsars younger than about $20\,000$~years (see \S\ref{sec_young}); older
pulsars with high space velocities can 
power bow-shock PWNe (see \S\ref{sec_bow}).

In the case of a dipole magnetic field, 
we find $k = 2M^2_\perp/3Ic^3$, where $M_\perp$ is the component
of the magnetic dipole moment orthogonal to the rotation axis.
We can thus calculate an equatorial surface magnetic field strength:
\begin{equation} 
B_p \equiv 3.2 \times 10^{19} (P \dot{P})^{1/2}~{\rm G}, 
\label{eqn_b} 
\end{equation} 
where $P$ is in seconds. Magnetic field strengths inferred from
Equation~(\ref{eqn_b}) range between $10^8$~G for recycled (or
``millisecond'') pulsars up to $>10^{15}$~G for ``magnetars''.
Most pulsars with prominent PWNe have inferred magnetic fields in the
range $1\times10^{12}$ to $5\times10^{13}$~G.

\subsubsection{Time Evolution of $\dot{E}$ and $P$}

A pulsar begins its life with an initial spin-down luminosity, $\dot{E}_0$.
If $n$ is constant, 
its spin-down luminosity then evolves with time, $t$, as
\citep[e.g.,][]{ps73}:
\begin{equation}
\dot{E} = \dot{E}_0 \left( 1 + \frac{t}{\tau_0}
\right)^{-\frac{(n+1)}{(n-1)}},
\label{eqn_edot_vs_t}
\end{equation}
where
\begin{equation}
\tau_0 \equiv \frac{P_0}{(n-1)\dot{P}_0} = \frac{2\tau_c}{n-1} - t
\label{eqn_tau0}
\end{equation} 
is the initial spin-down time scale of the pulsar. The pulsar
thus has roughly constant energy output until a time $\tau_0$,
beyond which $\dot{E} \propto t^{-(n+1)/(n-1)}$. The spin period
evolves similarly:
\begin{equation}
P = P_0 \left( 1 + \frac{t}{\tau_0} \right)^{\frac{1}{n-1}},
\label{eqn_p_vs_t}
\end{equation}
so that $P \approx P_0$ for $t \ll \tau_0$, but
at later times $P \propto t^{1/(n-1)}$.

\subsection{Radio and X-ray Emission from PWNe}
\label{sec_overall_emit}

As discussed in \S\ref{sec_young_pwn},
the resultant deposition of
energy with time generates a population of energetic electrons
and positrons, which in turn powers a synchrotron-emitting nebula.
Radio synchrotron emission is characterized by a power-law distribution
of flux, such that $S_\nu \propto \nu^\alpha$, where
$S_\nu$ is the observed flux density at frequency $\nu$, and $\alpha$
is the source's ``spectral index.''  At X-ray energies, the emission
is often described as a power-law distribution of photons, such that
$N_E \propto E^{-\Gamma}$, where $N_E$ is the number of photons emitted between
energies $E$ and $E+dE$, and $\Gamma \equiv 1 - \alpha$ is the ``photon
index.'' Typical indices for PWNe are $-0.3 \la \alpha \la 0$ in the radio
band, and $\Gamma \approx 2$ in the X-ray band.  This steepening of the
spectrum implies one or more spectral breaks between these two wavebands,
as discussed in \S\ref{sec_spectra}.

If the distance to a PWN is known,
the radio and X-ray luminosities, $L_R$ and $L_X$, respectively, can be
calculated over appropriate wavelength ranges. Typical ranges
are 100~MHz to 100~GHz for $L_R$ and 0.5--10~keV for $L_X$.
Observed values for $L_R$ and $L_X$ span many orders of magnitude,
but representative values might be $L_R \sim 10^{34}$~ergs~s$^{-1}$
and $L_X \sim 10^{35}$~ergs~s$^{-1}$.  The efficiency of
conversion of spin-down luminosity into synchrotron emission is defined
by efficiency factors $\eta_R \equiv L_R/\dot{E}$ and $\eta_X \equiv
L_X/\dot{E}$.
Typical values are $\eta_R \approx 10^{-4}$ and $\eta_X \approx
10^{-3}$ \citep{fs97,bt97}, although wide excursions from this are
observed. Note that if the synchrotron lifetime
of emitting particles is a significant fraction of the PWN age (as
is almost always the case at radio wavelengths, and sometimes also in
X-rays), then the PWN emission represents an integrated history of the
pulsar's spin down, and  $\eta_R$ and $\eta_X$ are not true instantaneous
efficiency factors.

%% file: sec3.tex
We now consider the phases of evolution which govern the overall
observational properties of PWNe.  The detailed theoretical
underpinning for these evolutionary phases is given in studies by
\cite{rc84}, \cite{che98,che05}, \cite{bcf01}, \cite{bbda03} and
van der Swaluw \etal\ (2004\nocite{vdk04}).

We defer a discussion of the details of how the wind is generated
to \S\ref{sec_young}, and here simply assume that the pulsar's
continuous energy injection ultimately results in an outflowing
wind which generates synchrotron emission.

\subsection{Expansion into Unshocked Ejecta}\label{sec_pwn_expand}

Since a pulsar is formed in a SN explosion, the star and its
PWN are initially surrounded by an expanding SNR.  The SNR blast wave 
at first moves outward freely at a speed $>(5-10)\times10^3$~\kms,
while asymmetry in the SN explosion gives the pulsar a
random space velocity of typical magnitude 400--500~\kms. At
early times the pulsar
is thus located near the SNR's center.

The pulsar is embedded in slowly moving unshocked ejecta from the
explosion and, since $t \ll \tau_0$, has constant energy output so
that $\dot{E} \approx \dot{E}_0$ (see Eqn.~[\ref{eqn_edot_vs_t}]). The
pulsar wind is highly over-pressured with respect to its environment,
and the PWN thus expands rapidly, moving supersonically and driving
a shock into the ejecta.  In the spherically symmetric case, the PWN
evolves as 
\citep{che77}:
\begin{eqnarray} 
R_{PWN} & \approx & 1.5  \dot{E}_0^{1/5} E_{SN}^{3/10} M_{ej}^{-1/2} t^{6/5}, \nonumber \\
       & = & 1.1~{\rm pc} \left( \frac{\dot{E}_0}{{\rm 10^{38}~ergs~s^{-1}}} \right )^{1/5}
          \left( \frac{E_{SN}}{{\rm 10^{51}~ergs}} \right)^{3/10}
          \left( \frac{M_{ej}}{{\rm 10~M_\odot}} \right)^{-1/2}
          \left( \frac{t}{{\rm 10^3~years}} \right)^{6/5},
\label{eqn_pwn_radius}
\end{eqnarray}
where $R_{PWN}$ is the radius of the PWN's forward shock at time $t$,
and $E_{SN}$ and $M_{ej}$ are the kinetic energy and ejected mass,
respectively, released in the SN.

\begin{figure}[htb!]
\centerline{\psfig{file=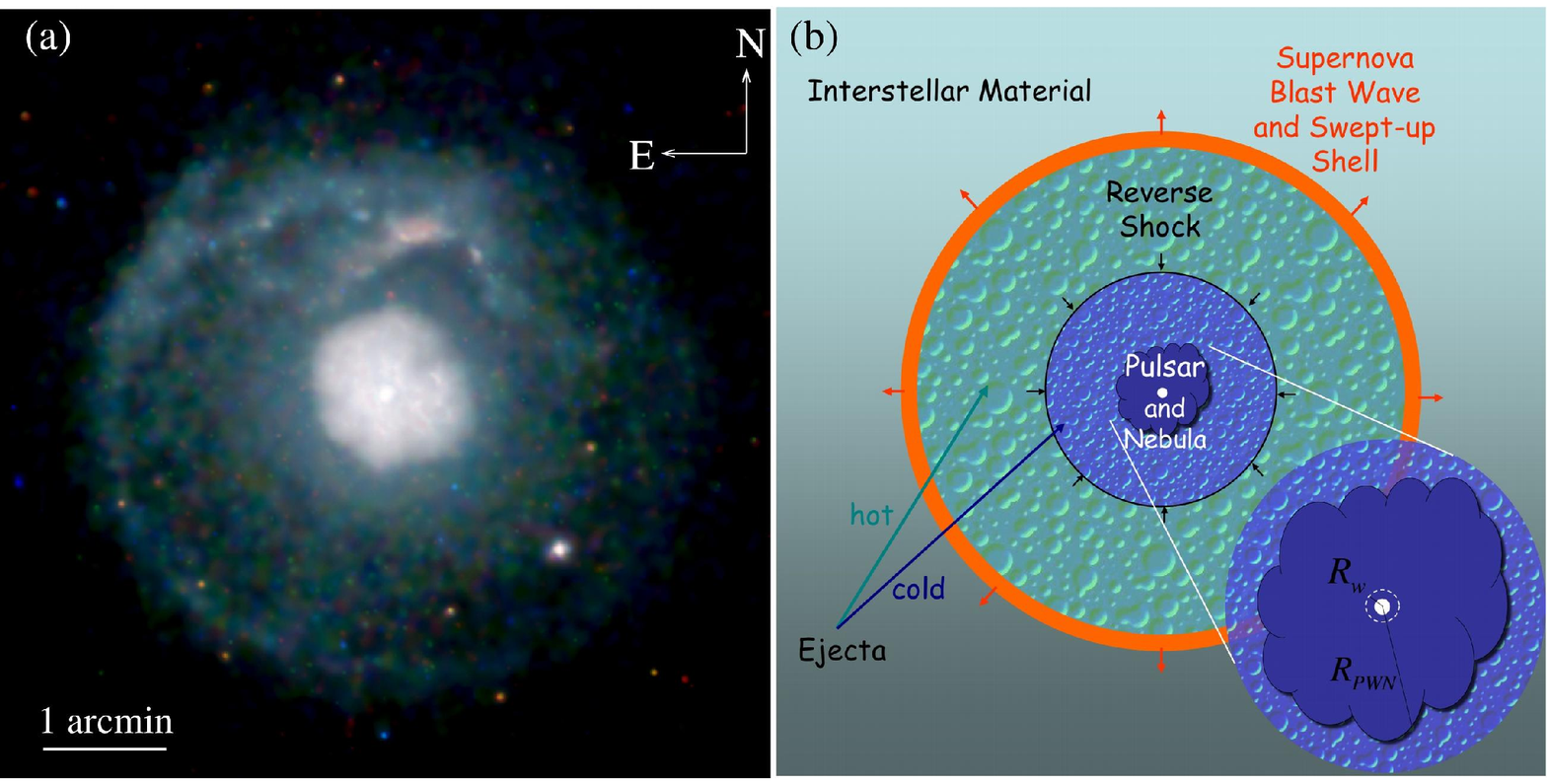,width=\textwidth}}
\caption{(a) A deep {\em Chandra}\ X-ray image of the composite
SNR~G21.5--0.9 \citep{ms05}. A circular SNR of diameter $\approx5'$
surrounds a symmetric PWN of diameter $\approx 1\farcm5$, with the young
pulsar J1833--1034 at the center \citep{gmga05,crg+05}. The central
location of the pulsar and PWN and the symmetric appearance of the PWN
and SNR both argue for a relatively unevolved system, in which the PWN
expands freely and symmetrically into the unshocked interior of the SNR.
(b) A schematic diagram of a composite SNR showing the swept-up ISM shell,
hot and cold ejecta separated by the reverse shock, and the central
pulsar and its nebula. The expanded PWN view shows the wind termination
shock. Note that this diagram does not correspond directly to G21.5--0.9, in
that a significant reverse shock has probably yet to form in this young SNR.}
\label{fig_g21}
\end{figure}

Since the PWN expansion velocity is steadily increasing, and the sound
speed in the relativistic fluid in the nebular interior is $c/\sqrt{3}$,
the PWN remains centered on the pulsar.  Observationally, we thus expect
to see a rapidly expanding SNR, with a reasonably symmetric PWN near its
center, and a young pulsar near the center of the PWN.  A good example
of system at this stage of evolution is the recently discovered pulsar
J1833--1034, which powers a bright X-ray and radio PWN, which in turn lies
at the center of the young SNR~G21.5--0.9
\citep[Fig.~\ref{fig_g21}(a);][]{ms05,gmga05,crg+05}.
This system is estimated to be $\sim1000$~years old.

\subsection{Interaction with the SNR Reverse Shock}

As the expanding SNR sweeps up significant mass from the ISM or
circumstellar medium,
it begins to evolve into the ``Sedov-Taylor'' phase, in which the
total energy is conserved and is partitioned equally between kinetic
and thermal contributions \citep[see][for a detailed discussion]{tm99a}.

The region of interaction between the SNR and its surroundings now
takes on a more complex structure, consisting of a forward shock
where ambient gas is compressed and heated, and a reverse shock where
ejecta are decelerated. The two shocks are separated by a contact
discontinuity at which instabilities can form.  The
reverse shock at first expands outward behind the
forward shock, but eventually moves inward.  In the absence
of a central pulsar or PWN, and assuming
that the SNR is expanding into a constant density medium
\citep[which, given the effects of progenitor 
mass loss by stellar winds, may not be the case; see][]{che05}, 
the reverse shock reaches the SNR center
in a time \citep{rc84}:
\begin{equation}
t_{Sedov} \approx 7 \left( \frac{M_{ej}}{10~M_\odot}\right)^{5/6}
\left( \frac{E_{SN}}{10^{51}~{\rm ergs}} \right)^{-1/2}
\left( \frac{n_0}{{\rm 1~cm}^{-3}} \right)^{-1/3}~{\rm kyr},
\label{eqn_sedov}
\end{equation}
where $n_0$ is the number density of ambient gas.
At this point the SNR interior is entirely filled
with shock-heated ejecta, and the SNR is in a fully self-similar
state which can be completely described by a small set of simple
equations \citep{cox72}.
The radius of the shell's forward shock now evolves
as $R_{SNR} \propto t^{2/5}$.

In the presence of a young pulsar, the inward moving SNR reverse
shock collides with the outward moving PWN forward shock after a
time $t_{coll} < t_{Sedov}$, typically a few thousand years (van
der Swaluw \etal\ 2001, Blondin \etal\ 2001\nocite{vagt01,bcf01}).
Even in the simplest case of a stationary pulsar, an isotropic wind
and a spherical SNR, the evolution is complex. The reverse shock
compresses the PWN by a large factor, which responds with an increase
in pressure and a sudden expansion.  The system reverberates several
times, resulting in oscillation of the nebula on a time scale of a
few thousand years, and a sudden increase in the nebular magnetic
field which serves to burn off the highest energy electrons
\citep{rc84,vagt01,bbda03}.  The crushing of the PWN produces
Rayleigh-Taylor instabilities, which can produce a chaotic, filamentary
structure and mixing of thermal and non-thermal material within the
PWN (Chevalier 1998, Blondin \etal\ 2001\nocite{che98,bcf01}).

\begin{figure}[b!]
\centerline{\psfig{file=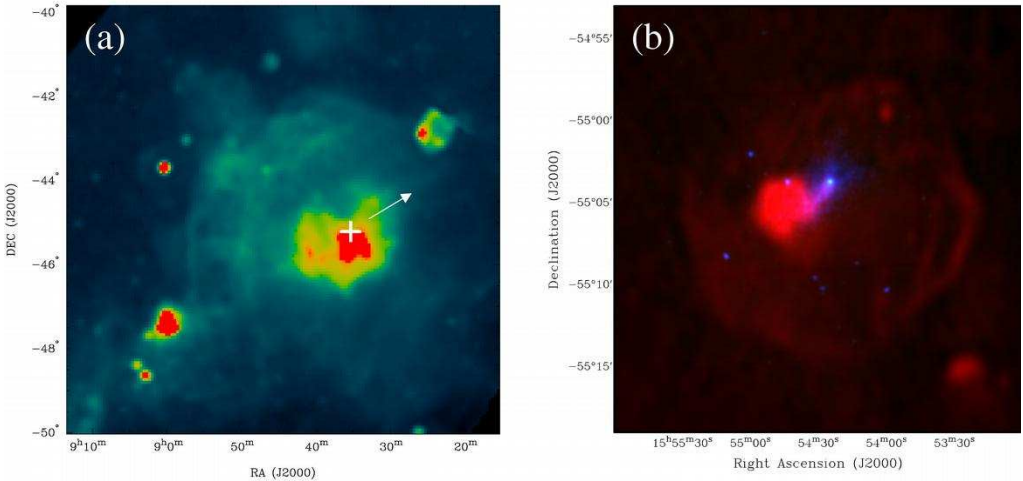,width=\textwidth,angle=90}}
\caption{(a) A 2.4-GHz Parkes map of the Vela SNR (G263.9--3.3),
\citep{dshj96}.
A limb-brightened shell and a central radio PWN can both be seen.
The cross indicates the location of the associated
pulsar B0833--45, while the white arrow
indicates its direction of motion \citep{dlrm03}.
The fact that the pulsar is neither at nor moving
away from the PWN's center indicates that a reverse shock
interaction has taken place.
(b) The composite SNR G327.1--1.1. An 843~MHz Molonglo image 
is shown in red \citep{wg96}, while a 0.2--12~keV {\em XMM-Newton}\ image
is in blue.
The radio morphology consists of a faint shell enclosing
a central PWN. 
The peak of X-ray emission indicates 
the likely position of an (as yet undetected) pulsar.
The offset between the X-ray
and radio nebulae indicates
that the radio nebula is a ``relic
PWN'' as discussed in \S\protect\ref{sec_evol_sedov}.  
The pulsar is likely to be still
moving subsonically through the SNR interior, generating
a new PWN as it moves away from its birthsite.}
\label{fig_vela_g327}
\end{figure}

In a more realistic situation, the pulsar's motion carries it away
from the SNR's center by the time the reverse shock collides with the
PWN. Furthermore, if the SNR has expanded asymmetrically, then the
reverse shock moves inward faster on some sides than on others. This
results in a complicated three-dimensional interaction, spread over
a significant time interval, during and after which the PWN can take on
a highly distorted morphology and be significantly displaced from the
pulsar position (Chevalier 1998, van der Swaluw
\etal\ 2004\nocite{che98,vdk04}).  An example of such a system is the
Vela~SNR, shown in Figure~\ref{fig_vela_g327}(a).

\subsection{A PWN inside a Sedov SNR}
\label{sec_evol_sedov}

Once the reverberations between the PWN and the SNR reverse shock
have faded, the pulsar can again power a steadily expanding 
bubble. However, the PWN now expands into hot, shocked, ejecta at subsonic
speeds. In the spherically symmetric case, there are two solutions,
depending on whether $t < \tau_0$  or $t > \tau_0$ (see
Eqn.~[\ref{eqn_tau0}]).  In the former case, $\dot{E}$
is approximately constant and the PWN radius evolves
as $R_{PWN} \propto t^{11/15}$ \citep{vagt01}, while for the latter
situation $\dot{E}$ is decaying, and (for $n=3$) we expect $R_{PWN}
\propto t^{3/10}$ \citep{rc84}.

At this point, the distance traveled by the pulsar from the explosion
site can become comparable to or even larger than the radius 
of an equivalent spherical PWN around a stationary pulsar.
The pulsar thus escapes from its original wind bubble, leaving
behind a ``relic PWN'', and generating a new, smaller PWN around
its current position (van der Swaluw
\etal\ 2004\nocite{vdk04}). Observationally, this appears
as a central, possibly distorted radio PWN, showing little corresponding
X-ray emission. The pulsar is to one side of or outside this region,
with a bridge of radio and X-ray emission linking it to the main
body of the nebula. An example 
is the PWN in the SNR~G327.1--1.1, 
shown in Figure~\ref{fig_vela_g327}(b).

The sound speed in the shocked ejecta drops as the pulsar moves
from the center to the edge of the SNR.  Eventually the pulsar's
space motion becomes supersonic, and it now drives a bow shock
through the SNR interior \citep{che98,vag98}.  The ram pressure
resulting from the pulsar's motion tightly confines the PWN, so
that the nebula's extent is small, $\la1$~pc.  Furthermore, the
pulsar wind is in pressure equilibrium with its surroundings, so
that the PWN no longer expands steadily with time.

\begin{figure}[b!]
\centerline{\psfig{file=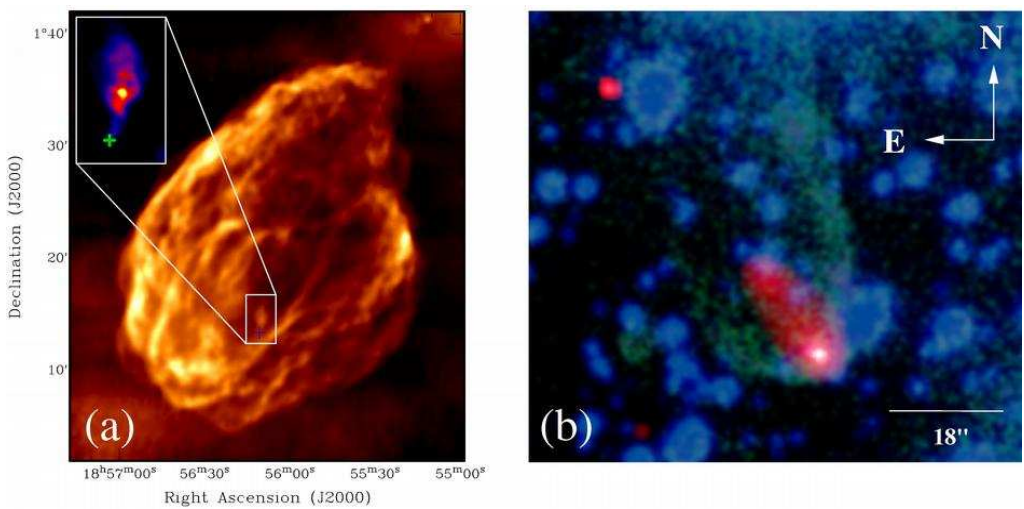,width=\textwidth,clip=}}
\caption{(a) The SNR W44 (G34.7--0.4). The main panel
shows a 1.4~GHz VLA image of the SNR
\citep{gdk+97}, while the inset shows
8.4~GHz VLA data on the region surrounding the associated young
pulsar B1853+01 \citep{fggd96}, whose position
is marked by a cross. The pulsar is nearing the edge of the SNR, and
now drives a small bow-shock PWN as a result of its supersonic motion.
(b) The recycled ``Black Widow'' pulsar (PSR~B1957+20) and its
bow shock \citep{sgk+03}. The green shows H$\alpha$ emission
imaged with the Anglo-Australian Observatory, while the red shows X-ray emission
observed with {\em Chandra}\ (the blue emission indicates
background stars). The pulsar is moving at a position angle
of $212^\circ$ (north through east).}
\label{fig_w44_b1957}
\end{figure}

For a SNR in the Sedov phase, the transition to a bow shock takes
place when the pulsar has moved 68\% of the distance between the
center and the forward shock of the SNR 
(van der Swaluw \etal\ 1998, 2003\nocite{vag98,vag+03}).  
The pulsar is now surrounded by a Mach
cone, and the PWN takes on a cometary appearance at X-ray and radio
wavelengths. An example of such a system is PSR~B1853+01 in the SNR~W44,
as shown in Figure~\ref{fig_w44_b1957}(a).

A pulsar will typically cross its SNR shell after $\sim40\,000$~years
(see Eqn.~[\ref{eqn_cross_sed}] in \S\ref{sec_bow}).
If the SNR is still in the Sedov phase, the bow shock has a Mach number
$\mathcal{M}$~$\approx 3.1$ at this point \citep{vag+03}.  The injection
of energy from the pulsar may brighten and re-energize the SNR shell
during its passage \citep{sfs89,vag02}.

\subsection{A Pulsar in Interstellar Gas}
\label{sec_evol_ism}

Once outside its SNR, a pulsar's motion is often highly supersonic
in interstellar gas. A bow-shock PWN results, with a
potentially large Mach number, $\mathcal{M}$~$\gg 1$.  

In cases where the pulsar propagates through neutral gas, the forward
shock driven by the PWN is visible, in the form of H$\alpha$ emission
produced by shock excitation and charge exchange (see \S\ref{sec_fwd}).
The shocked wind also produces synchrotron emission, resulting in
a bright head and cometary tail, both best seen in radio and X-rays
(see \S\ref{sec_term}). An example of an interstellar bow shock is
the structure seen around PSR~B1957+20, show in
Figure~\ref{fig_w44_b1957}(b).

As the pulsar now moves through the Galaxy, its $\dot{E}$ drops, and its
motion carries it away from the denser gas in the Galactic plane where
most neutron stars are born.  Eventually most pulsars will end up
with low spin-down luminosities in low density regions, where they may no
longer be moving supersonically and their energy output is insufficient
to power an observable synchrotron nebula. In this final stage, a pulsar
is surrounded by a static or slowly expanding cavity of relativistic
material with a radius $\gg1$~pc and confined by the thermal pressure
of the ISM \citep{bopr73,aro83}.  Deep searches have yet to detect such
``ghost nebulae'' \citep[e.g.,][]{gsf+00}.

An alternate evolutionary path may take place for old pulsars in
binary systems, which can eventually be spun up via accretion from
a companion.  This produces a recycled pulsar with a low value of
$\dot{P}$ but a very rapid spin period, $P\sim 1-10$~ms. Such pulsars
can have spin-down luminosities as high as $\dot{E} \approx
10^{34}-10^{35}$~ergs~s$^{-1}$, which is sufficient to generate
observable bow-shock nebulae, as shown in Figure~\ref{fig_w44_b1957}(b).

%% file: sec4.tex
\subsection{Pulsar Winds}\label{sec_young_pwn}

Despite more than 35 years of work on the formation of pulsar winds, there
are still large gaps in our understanding. The basic picture is that a
charge-filled magnetosphere surrounds the pulsar, and that particle
acceleration occurs in the collapse of charge-separated gaps either near
the pulsar polar caps or in outer regions that extend to the light cylinder
(i.e., to $R_{\rm LC} = c/\Omega$).
The maximum potential generated by the
rotating magnetic field has been calculated for the case of an aligned
rotator (i.e., with the magnetic and spin axes co-aligned) by \cite{gj69} as:
\begin{equation}
\Delta \Phi = \frac{B_p \Omega^2 R_{NS}^3}{2c} \approx 6 \times 10^{12}
\left(\frac{B_p}{10^{12}{\rm\ G}}\right) 
\left(\frac{R_{NS}}{10 {\rm\ km}}\right)^3 
\left(\frac{P}{1 {\rm\ s}}\right)^{-2} {\rm\ V},
\end{equation}
where $R_{NS}$ is the neutron star radius.
The associated particle current is
$\dot{N}_{GJ} = (\Omega^2 B_p R_{NS}^3)/Zec$
where $Ze$ is the ion charge. 
This current, although considerably modified in subsequent models, provides the
basis for the pulsar wind. In virtually all models, the wind leaving the
pulsar magnetosphere is dominated by the Poynting flux, $F_{E \times B}$,
with a much smaller
contribution from the particle energy flux, $F_{\rm particle}$. 
The
magnetization parameter, $\sigma$, is defined as:
\begin{equation}
\sigma \equiv \frac{F_{E \times B}}{F_{\rm particle}} = \frac{B^2}{4 \pi \rho 
\gamma c^2},
\label{eq_sigma}
\end{equation}
where $B$, $\rho$, and $\gamma$ are the  magnetic field,
mass density of particles, and Lorentz factor, in the wind, respectively.
As the wind flows from the pulsar light cylinder, 
typical values of $\sigma > 10^4$ are obtained.
However, models for the structure of the Crab Nebula 
\citep{rg74,kc84a}
require $\sigma \la 0.01$
just behind the termination shock (see \S\ref{sec_ts}), in 
order to meet flow and pressure boundary conditions at the 
outer edge of the PWN.
The high ratio of the synchrotron luminosity to the total spin-down power
also requires a particle-dominated wind \citep{kc84},
and implies $\gamma \sim 10^6$, a value
considerably higher than that expected in the freely expanding wind
\citep{aro02b}. Between the pulsar light cylinder
and the position of the wind termination shock the nature of the wind
must  thus change dramatically, although the mechanism  for
this transition is as yet unclear \citep[see][]{mel98,aro02b}.

The loss of electrons from the polar regions of the star represents a
net current that needs to be replenished to maintain charge neutrality.
This may occur through ion outflow in equatorial regions. As discussed in
\S\ref{sec_wisps}, ions may contribute to
nonthermal electron acceleration in the inner regions of the PWN.

\subsection{Observed Properties of Young PWNe}
\label{sec_young_obs}

\begin{figure}[t!]
\centerline{\psfig{file=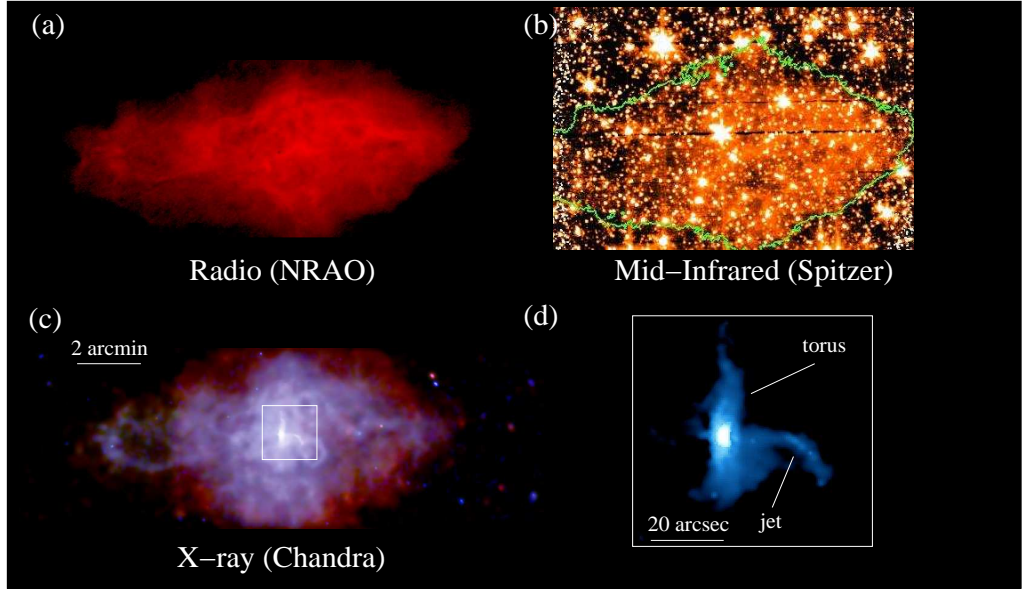,width=\textwidth}}
\caption{Images of the PWN 3C~58 (G130.7+3.1). 
(a) Radio synchrotron emission from
the confined wind, with filamentary structure \citep{ra88}. 
(b) Infrared synchrotron
emission with morphology similar to the radio nebula (whose outer contour
is shown in green).  (c) X-ray synchrotron emission (blue),
thermal emission (red) from shock-heated ejecta, 
and central torus/jet structure, shown expanded in (d)
\citep{shvm04}.  Images
are shown with north up and east to the left. The scale for the figures is
indicated by the 2 arcmin scale bar except for panel (d),
where the 20 arcsec scale bar applies. }
\label{fig_3c58}
\end{figure}

The deceleration of a  pulsar-driven wind as it expands into the
confines of cold, slowly expanding supernova ejecta produces a wind
termination shock, at which electron/positron pairs are accelerated
to ultrarelativistic energies (see \S\ref{sec_ts}).  As these
particles move through the wound-up magnetic field that comprises
the PWN, they produce synchrotron radiation from radio wavelengths
to beyond the X-ray band.  For a power-law electron spectrum,
the constant injection of particles plus a  finite synchrotron-emitting
lifetime lead to a spectral break at a frequency
\citep{gs65}:
\begin{equation}
\nu_b = 10^{21} \left(\frac{B_{\rm PWN}}{10^{-6}{\rm\ G}}\right)^{-3}
\left( \frac{t}{{\rm 10^3~years}}\right)^{-2}~{\rm\ Hz}, 
\label{eq_fbreak}
\end{equation} 
where $B_{\rm PWN}$ is the nebular magnetic field strength.
Particles radiating at frequencies beyond $\nu_b$ reach the outer
portions of the PWN in ever-diminishing numbers; most radiate their
energy before they are able to travel that far.  The result is that
the size of the PWN decreases with increasing frequency, as is
clearly observed in the Crab Nebula (Fig.~\ref{fig_crab}).  For
PWNe with low magnetic fields, the synchrotron loss times are longer
and there may not be a significant difference in size between the
radio and higher frequency bands (e.g., 3C~58 in Fig.~\ref{fig_3c58}).

The morphology of a young PWN is often elongated along the pulsar
spin axis due to the higher equatorial pressure associated with the
toroidal magnetic field \citep{bl92b,van03}.
This effect is seen clearly in many PWNe (e.g., Figs. 
\ref{fig_crab} \& \ref{fig_3c58})
and allows one to infer the likely projected orientation
of the pulsar.
As the nebula expands (see \S\ref{sec_pwn_expand}), 
Rayleigh-Taylor instabilities form
as the fast-moving relativistic fluid 
encounters and accelerates slower-moving unshocked
supernova ejecta.
These form dense, 
finger-like filamentary structures that
suffer photoionization from the surrounding synchrotron emission and
radiate recombination lines in the optical and
ultraviolet (UV) bands \citep[Fig.~\ref{fig_crab}(b);][]{hss96b}.  
The increased density compresses the magnetic field
around the filaments, causing enhanced synchrotron emission. One thus
observes radio structures that correspond to the optical/UV filaments.

At the core of the PWN lies the pulsar itself. As its free-flowing
equatorial wind encounters the more slowly-expanding nebula, a termination
shock is formed. Particles accelerated at the shock form a toroidal
structure, while some of the flow is collimated along the rotation
axis, possibly contributing to the formation of jet-like structures
\citep{bcku05}.  These structures generate synchrotron radiation that
is observed most readily in the X-ray band (Fig.~\ref{fig_crab}[d]),
although a toroid is also observed in optical images of the Crab
Nebula \citep{hss+95}.  The emission pattern from jets or ring-like
structures, as well as the larger scale geometry of the PWN, thus
provide an indication of the pulsar's orientation. As we discuss in
\S\ref{sec_wisps}, the emission structures in the post-shock and jet
regions provide direct insight on particle acceleration, magnetic
collimation and the magnetization properties of the winds in PWNe. In
addition, for pulsars whose proper motion is known, constraints on
mechanisms for producing this population's high-velocity birth kicks can
be derived based on the degree of alignment between the pulsar spin axis
and the direction of motion \citep{lcc01,nr04}.

\subsection{The Wind Termination Shock}\label{sec_ts}

The highly relativistic pulsar wind and its wound-up toroidal
magnetic field inflate an expanding bubble whose outer edge is
confined by the expanding shell of SN ejecta.  
As the wind is decelerated to match the boundary condition
imposed by the more slowly-expanding ambient material at the nebula
radius, a wind termination shock is formed at the radius, $R_w$,
at which the ram pressure of the wind is balanced by the internal
pressure of the PWN (Figure~\ref{fig_g21}[b]):
\begin{equation}
R_w = \sqrt{\dot E/(4 \pi \omega c \mathcal{P}_{\rm PWN})}, 
\label{eqn_ts}
\end{equation} 
where $\omega$ is the equivalent filling factor for an isotropic wind,
and $\mathcal{P}_{\rm PWN}$ is the total pressure in the shocked
nebular interior.
Upstream of the termination shock, the particles do not radiate, but flow
relativistically along with the frozen-in magnetic field. At the shock,
particles are thermalized and reaccelerated, 
producing synchrotron emission in the downstream flow. From
estimates of the field strength, the observed X-ray emission
implies Lorentz factors $\gamma \ga 10^6$ in the shock.

A reasonable pressure estimate can be obtained by
integrating the broad-band spectrum of the PWN, using standard
synchrotron theory \citep{gs65}, and assuming equipartition between
particles and the magnetic field. 
Typical $\mathcal{P}_{\rm PWN}$ and $\dot E$ values
yield termination shock radii of order 0.1~pc, implying an angular
size of several arcsec at distances of a few kpc. For the Crab
Nebula, the equipartition field is $B_{\rm PWN} \approx 300$~$\mu$G 
\citep{tri82}, and the pressure reaches equipartition at a radius of
$\sim (5 - 20) R_w$ \citep{kc84a}.
The associated spin-down luminosity, $\dot E = 5
\times 10^{38} {\rm\ ergs\ s}^{-1}$, yields $R_w \sim 4 \times
10^{17}$~cm, consistent with the position of the optical wisps and
the radius of the X-ray ring seen in 
Figure~\ref{fig_crab}(d).  Similar calculations indicate a much weaker
field, $B_{\rm PWN} \approx 80 \mu{\rm G}$ in 3C~58
\citep{gs92}, yielding a termination shock radius similar to the
Crab, $R_w \sim 6 \times 10^{17}$~cm, given the smaller $\dot E$
of the pulsar \citep{shm02}.
This lower field strength is also consistent with the fact
that the observed size of 3C~58 is similar in the radio
and X-ray bands (Fig.~\ref{fig_3c58}).

It must be noted that high resolution X-ray observations of 
3C~58 \citep{shm02}, G21.5--0.9 \citep{crg+05}, G292.0+1.8
\citep{hsb+01} and many other young PWNe and SNRs do not reveal directly
the ring-like emission that is observed just outside the termination shock
in the Crab pulsar, as seen
in Figure~\ref{fig_crab}(d). Rather, the compact
emission around the pulsar appears slightly extended (see 
Fig.~\ref{fig_3c58}[d]),
possibly originating from regions similar to the Crab
Nebula's torus, downstream 
from the termination shock.  However, the extent of such emission still 
provides a lower limit on $\mathcal{P}_{\rm PWN}$, 
as well as an indication of the
pulsar orientation.

\subsection{Formation of Tori, Jets and Wisps}
\label{sec_wisps}

The geometry implied by the X-ray morphology of the Crab Nebula
(see Fig.~\ref{fig_crab}[d]) is a tilted torus, with a jet of
material that flows along the toroid axis,
extending nearly 0.25~pc from the pulsar. A faint counter-jet
accompanies the structure, and the X-ray emission is
significantly enhanced along one edge of the torus.
Both effects are presumably the result of Doppler beaming of the
outflowing material, whereby the X-ray intensity, $I$, varies with viewing
angle  as \citep{ppp+87}:
\begin{equation} 
\frac{I}{I_0} = \left[\frac{\sqrt{1 - \beta^2}}{1 - \beta \cos\phi} 
\right]^{\Gamma +1} ,
\end{equation}
where $\beta c$ is the flow speed immediately
downstream of the termination shock,
$\phi$ is the angle of the flow to the line-of-sight, and
$I_0$ is the unbeamed intensity.

Similar geometric structures are observed in G54.1+0.3, for which
{\em Chandra}\ observations reveal a central point-like source
surrounded by an X-ray ring whose geometry suggests an inclination
angle of about $45^\circ$ \citep{lwa+02}.
The X-ray emission is brightest along the eastern limb of the ring.
If interpreted as the result of Doppler boosting, this implies a
post-shock flow velocity of $\sim 0.4c$. The ring is accompanied
by faint bipolar elongations aligned with the projected axis of the
ring, consistent with the notion that these are jets along the
pulsar rotation axis.
The total luminosity of these structures is similar to that of the
central ring. This is to be contrasted with the Crab and 3C~58
(Figs.~\ref{fig_crab} \& \ref{fig_3c58}), for which the torus outshines
the jets by a large factor.  Moreover, for G54.1+0.3 the brighter
portion of the outflow lies on the same side of the pulsar as the
brightest portion of the ring, which is inconsistent with 
Doppler boosting.
A similarly troubling observation is
that the brightness distribution around the inner ring of the Crab
also fails to show Doppler brightening consistent with that seen
in its surrounding torus; the brightness is reasonably uniform except
for some compact structures that vary in position and
brightness with time (see \S\ref{sec_struct}).

The formation of these jet/torus structures can be understood as
follows.  Outside the pulsar magnetosphere, the particle flow is
radial. The rotation of the pulsar forms an expanding toroidal
magnetic field for which the Poynting flux varies as $\sin^2\psi$,
where $\psi$ is the angle from the rotation axis. Conservation of
energy flux along flow lines leads to a latitude dependence of the
Lorentz factor of the wind of the form $\gamma = \gamma_0 + \gamma_m
\sin^2 \psi$ \citep{bk02} where $\gamma_0$ is the wind Lorentz factor just
outside the light cylinder \citep[$\gamma_0 \sim 10^2$ in standard
models for pulsar winds; e.g.,][]{dh82}, and $\gamma_m$ is the
maximum Lorentz factor of the preshock wind particles 
\citep[$\gamma_m \sim 10^6$ near the termination shock;][]{kc84a}.  From
Equation~(\ref{eq_sigma}), we see that this corresponds to a latitude
variation in the magnetization parameter also,
with $\sigma$ much larger at the equator
than at the poles.  This anisotropy results in the
toroidal structure of the downstream wind. Moreover, modeling of
the flow conditions across the shock shows that magnetic collimation
produces jet-like flows along the rotation axis \citep{kl04,bcku05}.
This collimation is highly dependent on the
magnetization of the wind. For $\sigma \ga 0.01$, magnetic hoop
stresses are sufficient to divert the toroidal flow back toward the
pulsar spin axis, collimating and accelerating the flow to 
speeds of 
$\sim 0.5c$ \citep{dab04}; smaller values of $\sigma$ lead to an increase in the
radius at which the flow is diverted. The launching points of the
jets observed in the Crab Nebula, which appear to form much closer to the
pulsar than the observed equatorial ring, apparently reflect such
variations in the value of $\sigma$. Near the poles $\sigma$ is
large, resulting in a small termination shock radius and strong
collimation, while near the equator $\sigma$ is much smaller and
the termination shock extends to larger radii \citep{bk02}.

Because $\gamma$ is smaller at high latitudes, models for the
brightness of the jet-like flows produced by the collimation process
fall short of what is observed.  However, kink instabilities in the
toroidal field may transform magnetic energy into particle energy,
thus accelerating particles and brightening the jet \citep{bcku05}.
Such instabilities limit the duration of the collimation \citep{beg98}.
Evidence for the effects of kink instabilities is seen in the
curved nature of many PWN jets, particularly for the Vela PWN where
the jet morphology is observed to change dramatically on timescales
of months \citep{ptks03}.  There appears to be a wide variation in
the fraction of $\dot{E}$ channeled into PWN jets, ranging from
$\sim2.5\times 10^{-5}$ in 3C~58 to nearly $10^{-3}$ for the PWN
powered by PSR~B1509--58, based on their synchrotron spectra
\citep{gak+02,shvm04}.  This apparently indicates considerable
differences in the efficiency with which additional acceleration
occurs along the jets.

The torus surrounding the Crab pulsar is characterized by the
presence of wisp-like structures (see Fig.~\ref{fig_crab}[d]), whose
position and brightness vary with time in the optical, infrared and X-ray
bands, and which emanate from the termination shock and move outward
through the torus  with inferred outflow speeds of $\sim0.5c$
\citep{hmb+02,msw+05}.  The exact nature of these structures is not fully
understood.  \cite{hmb+02} suggest that they are formed by synchrotron
instabilities.  However, the position of the arc-like structure
surrounding PSR~B1509--58 seems inconsistent with this hypothesis
given the much lower magnetic field in the PWN \citep{gak+02}.  An
alternative suggestion is that the wisps are the sites of compression
of the electron/positron 
pair plasma  on scales of the cyclotron gyration radius of
ions in the outflow, which can also explain the radius
of the X-ray arc seen around PSR~B1509--58 
\citep[][and references therein]{ga94,sa04}.

It is worth noting that VLA observations of the Crab Nebula
show variable radio structures very similar to the optical and X-ray wisps, 
indicating that acceleration of the associated particles must be occurring in
the same region as for the X-ray-emitting population \citep{bhfb04}.

\subsection{Filamentary and Compact Structures in PWNe}
\label{sec_struct}

In the Crab Nebula, an
extensive network of filaments is observed in H$\alpha$, [O\,{\sc
iii}] and other optical lines, surrounding the nonthermal optical
emission from the PWN (Fig.~\ref{fig_crab}[b]).
The detailed morphology and ionization structure of these filaments
indicate that they form from Rayleigh-Taylor instabilities as the
expanding relativistic bubble sweeps up and accelerates slower moving ejecta
\citep{hss96b}.  Magnetohydrodynamic (MHD) simulations support this
picture, indicating that 60\%--75\% of the swept-up mass can be
concentrated in such filaments \citep{jun98,bab+04}. As the expanding
PWN encounters these filaments, compression increases the density
and magnetic field strength, forming sheaths of enhanced synchrotron
emission observed as a corresponding shell of radio filaments
\citep{rey88b}. X-ray observations reveal no such filaments in the
Crab Nebula \citep{wht+00}.  This is presumably
because the higher energy electrons required to produce the X-ray
emission suffer synchrotron losses before reaching the outer regions
of the PWN, consistent with the smaller extent of the nebula in
X-rays than in the radio.

A different picture is presented by X-ray observations of 3C~58,
which reveal a complex of loop-like filaments most prominent near the
central regions of the PWN, but evident throughout the nebula 
\citep[Fig.~\ref{fig_3c58}(c);][]{shvm04}.  
These nonthermal X-ray structures
align with filaments observed in the radio band \citep{ra88}.
Optical observations of 3C~58 also reveal faint filaments \citep{van78},
whose origin is presumably similar to those in the Crab Nebula. However,
a detailed X-ray/optical comparison shows that most
of the X-ray filaments do not have optical counterparts.
While comparisons with deeper optical images are clearly needed,
the fact that many of the X-ray features without optical counterparts
are brighter than average suggests that these may actually
arise from a different mechanism. \cite{shvm04} propose that the
bulk of the discrete structures seen in the X-ray and radio images
of 3C~58 are magnetic loops torn from the toroidal field by kink
instabilities. In the inner nebula, the loop sizes are similar to the
size of the termination shock radius ($\sim 0.1$~pc), as suggested by
\cite{beg98}.  As the structures expand, they enlarge slightly as a
consequence of decreasing pressure.

There is also considerable loop-like filamentary structure evident
in high resolution X-ray images of the Crab Nebula
\citep[Fig.~\ref{fig_crab}(d);][]{wht+00}.  These filaments appear to
wrap around the torus, perpendicular to the toroidal plane, and may
be signatures of kink instabilities in the termination shock region.

In some PWNe, compact knot-like structures
are observed close to the pulsar,  which dissipate and reappear on
timescales of order months \citep{hmb+02,msw+05}.  Examples can be seen
for the Crab in Figure~\ref{fig_crab}(d), and similar structures are
also seen for PSR~B1509--58 \citep{gak+02}.
Some of these features appear in projection inside the
termination shock region. However, it is believed that they actually
correspond to unstable, quasi-stationary shocks in the region just
outside the termination shock, at high latitudes where the shock
radius is small due to larger values of $\sigma$ \citep[e.g.,][]{kl04}.

\subsection{PWN Spectra}\label{sec_spectra}

As noted in \S\ref{sec_overall_emit}, the spectra of PWNe are characterized by
a flat power law index at radio wavelengths ($\alpha \approx -0.3$)
and a considerably steeper index
in X-rays ($\Gamma \approx 2)$.  The nature of this spectral
steepening is not understood.  Simple assumptions of a power-law
particle spectrum injected by the pulsar would predict a power-law
synchrotron spectrum, with a break associated with the aging of the
particles (see Eqn.~[\ref{eq_fbreak}]); the expected increase in
spectral index is $\Delta \alpha = 0.5$, which is smaller than what is typically
observed \citep{wspb97}.  
Moreover, for many PWNe a change in spectral index is
inferred at low frequencies that would imply unrealistically high
magnetic fields \citep[e.g.,][]{gs92}. 
Relic breaks in the spectrum can be produced by a
rapid decline in the pulsar output over time, and these breaks propagate to
lower frequencies as the PWN ages \citep{ps73}. 
The inherent spectrum of the injected
particles, which may deviate from a simple power law, as well as
modifications from discrete acceleration sites, all contribute to a
complicated integrated spectrum. As a result, the interpretation of
spectral steepening as being due to synchrotron losses can
lead to drastically wrong conclusions about PWN properties.

At frequencies for which the synchrotron lifetime 
is shorter than the flow time to the edge of the PWN, one expects a
steepening of the spectrum with radius. Radial steepening is indeed observed
in the X-ray spectra of the Crab Nebula, 3C~58 and G21.5--0.9
\citep{scs+00,wag+01,shvm04},
but the spectra steepen rather
uniformly with radius whereas generalizations of the \cite{kc84}
model predict a much more rapid steepening near the outer regions
\citep{rey03}. Some
mixing of electrons of different ages at each radius seems to be required,
perhaps due to diffusion processes in the PWN.

%% file: sec5.tex
Pulsars are born with high
space velocities, typically $V_{\rm PSR} =$~400--500~\kms, but
for some sources exceeding 1000~\kms.
These high velocities are almost certainly the result of kicks
given to the star during or shortly after core collapse \citep{lai04}.
Young pulsars thus have the highest velocities of
any stellar population, and many have sufficient speeds to eventually
escape the Galaxy.

As discussed in \S\ref{sec_evol_sedov}, a pulsar's ballistic motion allows
it to eventually escape its original PWN, and to propagate through the
shocked ejecta in the SNR interior.  
At first the pulsar's motion will be subsonic in this hot gas, but by the 
time the pulsar nears the edges of the SNR, the sound speed drops
sufficiently for the pulsar's motion to be supersonic. 
In the simplest situation of a spherical SNR in the Sedov phase, expanding
into a uniform medium, this transition occurs at half the crossing time
(given in Eqn.~[\ref{eqn_cross_sed}] below), at which point the pulsar
has traveled 68\% of the distance from the center of the SNR to its edge
(van der Swaluw \etal\ 2004\nocite{vdk04}).
This simple result is 
independent of $V_{\rm PSR}$, $n_0$ or $E_{SN}$.
The pulsar's supersonic motion now produces a PWN
with a bow shock morphology.

Since the SNR is decelerating,
the pulsar ultimately penetrates and then escapes the shell.
A pulsar moving at $V_{\rm PSR} \ga 650$~\kms\
will escape while the SNR is 
still in the Sedov phase, after a time \citep{vag+03}:
\begin{equation}
t_{cross} = 44~\left( \frac{E_{SN}}{10^{51}~{\rm ergs}}\right)^{1/3}
\left( \frac{n_0}{{\rm 1~cm}^{-3}} \right)^{-1/3}
\left( \frac{V_{\rm PSR}}{500~{\rm km~s}^{-1}}\right)^{-5/3}~{\rm kyr}.
\label{eqn_cross_sed}
\end{equation}
At times $t > t_{cross}$, a pulsar proceeds to move through the ambient ISM.

The speed of sound in interstellar gas is a a function of temperature:
typical values are approximately 1, 10 and 100~\kms\ for the cold, warm and
hot components of the ISM, respectively.  Thus except in
the case of a particularly slow moving pulsar moving through coronal gas,
a pulsar will move supersonically and drive a bow shock through the ISM.

\subsection{Theoretical Expectations}
\label{sec_theory}

The pulsar wind in a bow shock is decelerated at a
termination shock, just as for younger PWNe 
(see \S\ref{sec_ts}).
However, the external source of pressure balance is now ram pressure
from the pulsar's motion, rather than the internal pressure of the
shocked wind. Furthermore, since ram pressure is not isotropic, the termination
shock radius varies as a function of angle with respect to the pulsar's
velocity vector. In the direction of the star's motion, the termination shock radius
is referred to as the ``stand-off distance'', $R_{w0}$, and is defined by
(cf.~Eqn.~[\ref{eqn_ts}]):
\begin{equation}
\frac{\dot{E}}{4\pi \omega R_{w0}^2 c} = \rho_0 V_{\rm PSR}^2,
\label{eqn_balance}
\end{equation}
where $\rho_0$ is the ambient density.
If the wind is isotropic and $\omega = 1$,
then at a polar angle $\theta$ with respect to the bow shock's symmetry
axis, the analytic solution for the
termination shock radius as a function of position is \citep{wil96}:
\begin{equation}
R_w(\theta) = R_{w0} \csc \theta \sqrt{3(1-\theta \cot \theta)}.
\label{eqn_wilkin}
\end{equation}
It is important to note that the above solution assumes an efficiently
cooled thin-layer shock, in contrast to the double shock expected for
pulsar bow shocks.  Full hydrodynamic and MHD simulations show that
Equation~(\ref{eqn_wilkin}) is a reasonable approximation
in regions near the apex ($\theta \la \frac{\pi}{2}$),
but performs more poorly further downstream \citep{buc02a,vag+03}.

\begin{figure}[t!]
\vspace{-11mm}
\centerline{\psfig{file=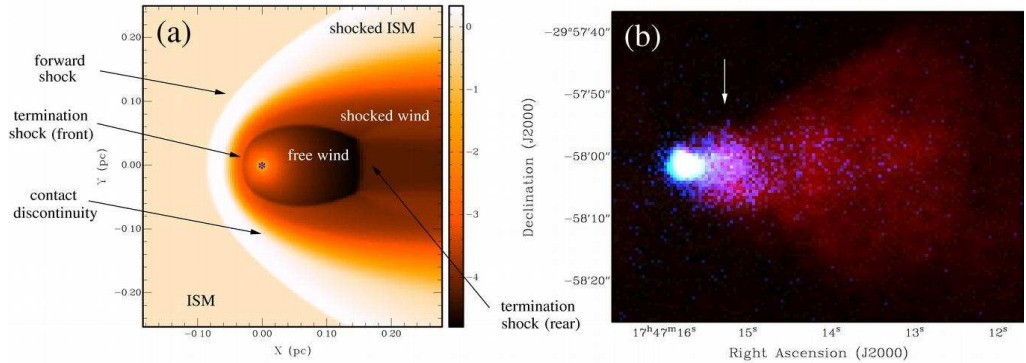,width=\textwidth,clip=}}
\vspace{-7mm}
\caption{(a) A hydrodynamic simulation of a pulsar bow shock, adapted from
Gaensler \etal\ (2004\protect\nocite{gvc+04}).  The pulsar, whose position
is marked with an asterisk, is moving from right to left with a Mach
number $\mathcal{M}$~$= 60$.  The intensity in the image and the scale-bar
indicate density, in units of $\log_{10} (\rho_0/ 10^{-24}~{\rm g~cm}^{-3})$.
(b) {\em Chandra}\ X-ray (blue) and VLA radio (red) images of 
G359.23--0.82
(``the Mouse''), the bow shock associated
with PSR~J1747--2958 \citep{gvc+04}.
The white arrow marks a bright compact region of X-ray
emission behind the apex, which possibly corresponds to the
surface of the termination shock.}
\label{fig_hydro_mouse}
\end{figure}

A result of such simulations is shown in Figure~\ref{fig_hydro_mouse}(a). The
double-shock structure is clearly apparent, consisting of a forward
shock where the ISM is heated, plus the termination shock where the
pulsar's wind decelerates. As expected, the termination shock is
not of uniform radius around the pulsar: specifically, for low Mach
numbers, $\mathcal{M}$~$\sim 1-3$ (as may be appropriate for pulsars
traveling supersonically inside their SNRs; see \S\ref{sec_evol_sedov}
and van der Swaluw \etal\ 2004\nocite{vdk04}), the ratio
of termination shock radii between polar angles $\theta = \pi$ and
$\theta = 0$ is approximately $\mathcal{M}$ \citep{buc02a,vag+03}, but
for $\mathcal{M}$~$\gg 1$ (typical of bow shocks in the ambient ISM;
see \S\ref{sec_evol_ism}), this ratio approaches a limit of $\sim5-6$
\citep{gvc+04,bad05}.

\subsection{Observations: Forward Shock}
\label{sec_fwd}

If a pulsar drives a bow shock through neutral gas, then collisional
excitation and charge exchange occur at the forward shock, generating
optical emission in the Balmer lines
\citep{bb01,buc02b}.  Indeed several pulsar bow shocks have been
identified in the 656-nm H$\alpha$ line, 
associated with (in  order of
discovery): B1957+20 \citep[Fig.~\ref{fig_w44_b1957}(b);][]{kh88}, B2224+65
\citep[``the Guitar'';][]{crl93}, J0437--4715 \citep{bbm+95},
RX~J1856.5--3754 \citep{vk01},
B0740--28 \citep{jsg02} and J2124--3358 \citep{gjs02}.

If the distance to the system is known, $R_{w0}$ can be
directly measured, provided that one adopts a scaling
factor of $\sim0.4-0.6$ to translate between the observed radius of the
forward shock to that of the termination shock \citep{buc02a,vag+03}.
If $\dot{E}$ and $V_{\rm PSR}$ have been measured, Equation~(\ref{eqn_balance})
can then be applied to yield $\rho_0$. This is an approximation,
since $\omega=1$ is usually assumed, and
because of the unknown inclination of the pulsar's motion to the line of
sight,\footnote{Note that the correction factor
for motion inclined to the line of sight cannot be derived through
simple trigonometry (Gaensler \etal\ 2002b\nocite{gjs02}).}
but certainly suggests ambient number densities $\sim0.1$~cm$^{-3}$, as expected
for warm neutral ambient gas
(Chatterjee \& Cordes 2002, Gaensler \etal\ 2002b\nocite{cc02,gjs02}). For
pulsars for which $V_{\rm PSR}$
is not known, one can write: 
\begin{equation} 
\rho_0 V_{\rm PSR}^2 = \gamma_1{\mathcal{M}}^2\mathcal{P}_{\rm ISM}, 
\label{eqn_mach} 
\end{equation} 
where $\gamma_1 = 5/3$
and $\mathcal{P}_{\rm ISM}$ are the adiabatic coefficient and pressure
of the ISM, respectively. Estimates for
$\mathcal{P}_{\rm ISM}$
and $\omega$ in Equations (\ref{eqn_balance}) \& (\ref{eqn_mach}) then 
yield $\mathcal{M}$.

For the bow shocks associated with RX~J1856.5--3754, PSR~J0437--4715
and PSR~B1957+20
(Fig.~\ref{fig_w44_b1957}[b]),
the shape of the forward shock is a good
match to the solution predicted by Equation~(\ref{eqn_wilkin})
and simulated in Figure~\ref{fig_hydro_mouse}(a).  However, the optical
emission round PSRs~B2224+65, J2124--3358 and B0740--28 all show
strong deviations from the expected shape, in that there are abrupt
kinks and inflection points in the H$\alpha$ profile
(Jones \etal\ 2002, Chatterjee \& Cordes 2002\nocite{jsg02,cc02}).
Furthermore, in the case of PSR~J2124--3358
there is an apparent rotational offset between the symmetry axis of
the bow shock and the velocity vector of the pulsar (Gaensler
\etal\ 2002b\nocite{gjs02}).
These systems imply the presence of some combination of anisotropies
in the pulsar wind (as are observed in young pulsars;
see \S\ref{sec_wisps}), 
gradients and fluctuations in the density of the ISM, or a
bulk velocity of ambient gas with respect to the pulsar's local standard
of rest.  Extensions of Equation~(\ref{eqn_wilkin}) to account for these
effects have been presented by \cite{ban93} and by 
\cite{wil00}, and have been applied to interpret the morphology
of PSR~J2124--3358 by Gaensler \etal\ (2002b\nocite{gjs02}).

\subsection{Observations: Termination Shock}
\label{sec_term}

Just as for the PWNe around the youngest pulsars discussed in \S\ref{sec_ts},
particles in the pulsar wind inside a bow shock will be accelerated
at the termination shock, producing non-thermal synchrotron emission
that can be potentially observed in the radio and X-ray bands.
Indeed cometary radio and X-ray PWNe aligned with the direction of
motion has been now identified around many pulsars, convincing
examples of which include PSRs~B1853+01
\citep[Fig.~\ref{fig_w44_b1957}(a);][]{fggd96,pks02}, B1957+20
\citep[Fig.~\ref{fig_w44_b1957}(b);][]{sgk+03} and B1757--24 \citep[``the
Duck'';][]{fk91,kggl01}.

The most spectacular example of this class is G359.23--0.82,
the X-ray/radio bow
shock powered by PSR~J1747--2958 \citep[``the Mouse'';][]{yb87,gvc+04},
multi-wavelength observations of
which are shown in Figure~\ref{fig_hydro_mouse}(b).  The extent of
the radio trail in this system is larger than in
X-rays, reflecting the difference in synchrotron lifetimes between these
bands (cf.\ \S\ref{sec_young_obs}).

The X-ray morphology of the Mouse in Figure~\ref{fig_hydro_mouse}(b)
appears to consist of two components: a bright compact region extending
$\sim0.2$~pc from the pulsar, superimposed on a larger fainter component
extending $\sim 1$~pc from the pulsar.  Comparison with the hydrodynamic
simulation in Figure~\ref{fig_hydro_mouse}(a) suggests that the bright component
of the X-ray emission corresponds to the surface of the wind termination
shock, while fainter, more extended X-ray emission originates from the
shocked wind \citep{gvc+04}.  In this identification, the bright X-ray
component of this bow shock corresponds to the inner toroidal
rings discussed in \S\ref{sec_wisps}, 
but elongated due to the pulsar motion. This
interpretation is supported by the fact that the extent of the termination
shock region along the symmetry axis (in units of $R_{w0}$) should
be a function of Mach number, as was discussed in \S\ref{sec_theory}.
Indeed  for the Mouse, with ${\mathcal{M}} \sim 60$
\citep{gvc+04}, this bright region is about twice as long (relative to
$R_{w0}$) as for the pulsar bow shock seen inside the SNR~IC~443
\citep{ocw+01},
which must have ${\mathcal{M}}
\ga 3$ since it is moving through shocked gas in the SNR interior
\citep{vag+03}.

In PWNe for PSRs~B1757--24 \citep{kggl01} and B1957+20
\citep[Fig.~\ref{fig_w44_b1957}(b)][]{sgk+03}, only a short
($\sim0.1$~pc) narrow X-ray trail is apparent.  Such features have been
interpreted as a rapid back-flow or nozzle, which transports
particles downstream \citep{wlb93,kggl01}.  However,
Figure~\ref{fig_hydro_mouse} suggests
that the short trail seen behind PSRs~B1757--24 and B1957+20 is the
surface of the termination shock, and that 
emission from the post-shock wind further downstream is too faint
to see \citep{gva04b,gvc+04}.  Deeper X-ray observations are required to
test this possibility.

Hydrodynamic models predict that the pulsar's motion should divide the
post-shock emitting region of a bow shock into two distinct zones: a
highly magnetized broad tail originating from material shocked
at $\theta \la \frac{\pi}{2}$,
plus a more weakly magnetized, narrow, collimated
tail, produced by material flowing along the axis
$\theta \approx \pi$ \citep{buc02a,rcl05}.  These two structures 
are both apparent in the Mouse \citep{gvc+04}.
Through relativistic MHD simulation,
this and other issues related to the structure of the post-shock
flow are now being explored (e.g., Bucciantini \etal\ 2005\nocite{bad05}).

%% file: sec6.tex
\subsection{Pulsars and PWNe in Very Young SNRs}

The youngest known Galactic PWNe are the Crab Nebula and 3C~58,
powered by pulsars thought to correspond to the
SNe of 1054~CE and 1181~CE, respectively.\footnote{The association of
3C~58 with 1181~CE is not completely secure; see 
\cite{sg02b} and \cite{che05}, and
references therein.} It would be of great interest to identify
PWNe at earlier evolutionary stages.

SN~1987A formed a neutron star, but deep searches have failed
to detect a central object, down to a luminosity
$\la10^{34}$~ergs~s$^{-1}$ \citep[e.g.,][]{gcc+05}.
This is well below the luminosity of the Crab Nebula, and may indicate
that the central neutron star has collapsed further into a black hole,
accretes from fall-back material, or does not generate
a wind \citep[e.g.,][]{fcp99}.

Searches for PWNe in other extragalactic SNe and SNRs have generally not
produced any convincing candidates \citep{rf87b,bb05}. However, recent high
resolution radio images have revealed the gradual turn-on of a central
flat-spectrum radio nebula in SN~1986J, 
which may correspond to emission from a very young PWN
\citep{bbr04}.  New wide-field radio and X-ray images of other galaxies
may lead to further identification of young PWNe,
while optical and UV spectroscopy of recent SNe may identify
PWNe through emission lines which broaden with time as gas is swept up
by the expanding pulsar wind \citep{cf92}.

\subsection{Winds from Highly Magnetized Neutron Stars}

A growing population of neutron stars have surface magnetic fields (inferred
via Eqn.~[\ref{eqn_b}]) above the quantum critical limit
of $B_p = 4.4\times10^{13}$~G. The properties of these
stars indicate that they are comprised of two apparently distinct
populations: the high-field radio
pulsars \citep[][and references therein]{msk+03},
and the exotic magnetars \citep{wt04}. The winds and PWNe of these sources
potentially provide a view of different spin-down processes than those
seen in normal pulsars.

Most of the high-field radio pulsars have $\dot{E} \approx 10^{32}
- 10^{34}$~ergs~s$^{-1}$. For typical efficiency factors $\eta_R,
\eta_X \la 10^{-3}$, this implies PWNe too faint to be detectable.
However, the very young pulsars J1846--0258 ($B_p = 4.9\times10^{13}$~G)
and J1119--6127 ($B_p = 4.1\times10^{14}$~G) have high spin-down
luminosities ($\dot{E} > 10^{36}$~ergs~s$^{-1}$) and are near the
centers of SNRs.  In both cases PWNe are detected, although with
very different properties.  PSR~J1846--0258 puts a large fraction
($\eta_X \sim 0.2$) of its spin-down power into a luminous X-ray
PWN $\sim2$~pc in extent \citep{hcg03}, while PSR~J1119--6127 powers
an under-luminous ($\eta_X \sim2\times10^{-4}$) and small ($\sim0.2$~pc)
X-ray nebula \citep{gs03b}.  Clearly PWN properties are dominated
by factors such as age, environment and evolutionary state (see
\S\ref{sec_evol}) rather than the associated pulsar's surface
magnetic field.

Magnetars also spin down, albeit in some cases not
smoothly \citep{wkg+02}.  Just as for radio pulsars, this rotational
energy output is thought to go into a relativistic wind, but
traditional PWNe have not been detected around magnetars,\footnote{The
magnetar SGR~1806--20 was originally presumed to power the radio
nebula G10.0--0.3, but a revision in the position of this neutron star now
makes this unlikely \citep{hkc+99}.}
presumably because $\dot{E} \la 10^{34}$~ergs~s$^{-1}$ for all these
sources.  Magnetars likely experience an enhanced torque
over the dipole spin-down presumed to act in radio pulsars, as a
result of either Alfv\'en waves and outflowing relativistic particles
driven by seismic activity, or by a large-scale twist of the external
magnetic field \citep{hck99,tlk02}.  Under either circumstance,
the spin-down behavior deviates from that described in Equations
(\ref{eqn_tau}) to (\ref{eqn_p_vs_t}) \citep[e.g.,][]{tdw+00}.

A transient radio PWN was proposed to account for the short-lived
radio nebula seen in 1998 following the giant flare from the
magnetar SGR~1900+14 \citep{fkb99}. However, recent
observations of a radio nebula in the aftermath of a giant flare
from SGR~1806--20 suggest that the synchrotron emission from these
nebulae are powered by ejected baryonic material,
so that these sources are more analogous to SNRs than to
PWNe \citep{gkg+05}.\footnote{A possibly transient radio source has
also recently been identified coincident with the flaring magnetar
XTE~J1810--197, but the nature of this source is as yet unclear
\citep{hgb+05}.}

\subsection{TeV Observations of PWNe}

The Crab Nebula is a well-known source of TeV gamma-rays \citep[][and
references therein]{wcf+89}.  This emission is well explained as
inverse Compton (IC) emission, the relativistic particles in the
shocked wind acting as scattering centers for the synchrotron photons
that they themselves emit at lower energies \citep{aa96}.  Under
this interpretation, the emitted spectrum can be modeled to provide
the mean and spatial distribution of the nebular magnetic field strength,
and hence the PWN's particle content, the
time-averaged injection rate of particles, and an
independent estimate of the magnetization parameter, $\sigma$
\citep[e.g.,][]{dhm+96}. The estimated values of $B_{\rm PWN}$ and
$\sigma$ are in good agreement with those derived from the MHD model
of \cite{kc84a}.

\begin{figure}[t!]
\centerline{\psfig{file=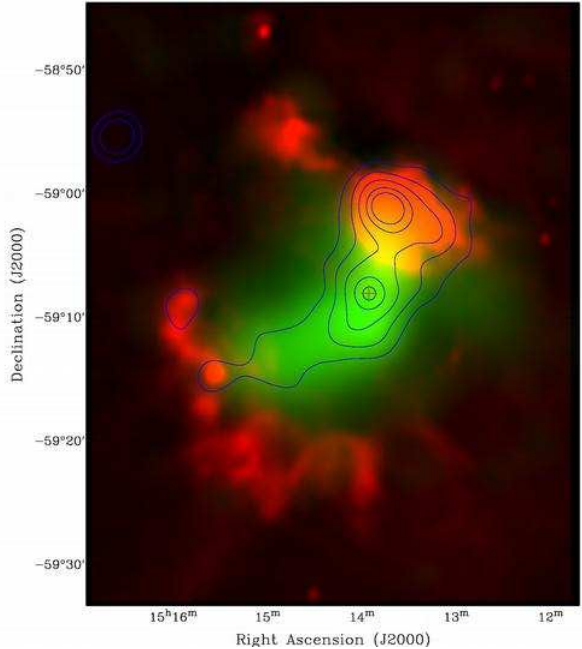,width=0.7\textwidth,angle=270}}
\caption{Multi-wavelength images of the PWN 
powered by the young pulsar B1509--58.  {\em ROSAT}\ PSPC data
(in blue contours, at levels of 5\%, 10\%, 20\%, 40\% and 60\% of the
peak) show the extent of the X-ray PWN \citep{tmc+96}, while 843~MHz
Molonglo data (in red) correspond to the surrounding SNR G320.4--1.2
\citep{wg96}.  TeV emission from HESS is shown in green \citep{aaa+05b}.  
The cross marks the position of PSR~B1509--58}
\label{fig_g320_tev}
\end{figure}

The new generation of ground-based \v{C}erenkov detectors (most
notably the High Energy Stereoscopic System, HESS) have now begun
to detect other PWNe in the TeV band.  These detections
indicate that acceleration of particles to considerable energies
must have occurred, and provide estimates of the nebular magnetic
field strength which can be used in modeling and interpreting the
other nebular structures discussed in \S\ref{sec_young}.  The much
lower synchrotron luminosities of these other sources compared to
the Crab Nebula implies that the seed photons for IC scattering are in
these cases primarily external, originating from a combination of
the cosmic microwave background and a local contribution from dust
and starlight.  Recent TeV detections of PWNe by HESS include
G0.9+0.1 and G320.4--1.2 / PSR~B1509--58 \citep{aaa+05a,aaa+05b}.
As shown in Figure~\ref{fig_g320_tev}, the latter is spatially
resolved by HESS and has a TeV morphology which is a good match to
the X-ray synchrotron nebula.  Such observations can potentially
provide direct measurements of spatial variations in the magnetic
fields of PWNe.

\subsection{Pulsar Winds in Binary Systems}

The recently discovered dual-line double pulsar PSR~J0737--3039 consists
of a 23-ms pulsar (``A'') and a 2.8-s pulsar (``B'') in a 2.4-hour orbit,
viewed virtually edge-on \citep{lbk+04}.  This system is proving to be
a remarkable new probe of pulsars and their winds, providing information
at much closer separations to the pulsar than is possible for the sources
discussed in \S\ref{sec_young} \& \S\ref{sec_bow}.

The line of sight to pulsar~A passes within $\approx0.01$~light-seconds
of pulsar~B, well within the unperturbed light cylinder radius of the
slower pulsar.  For $\approx30$~s at conjunction, the pulse-averaged
radio emission from A is modulated with a complicated time-dependence,
showing intermittent periodicities at both 50\% and 100\% of the
rotational period of pulsar~B \citep{mll+04}.  Detailed modeling shows
that this behavior can be interpreted as synchrotron absorption from
a relativistic plasma confined by a dipolar magnetic field, providing
direct evidence for the field geometry commonly adopted in pulsar
electrodynamics \citep{rg05,lt05}.

Drifting sub-pulses in the pulsed emission from pulsar~B are also
observed, with fluctuations at the beat frequency between the
periods of A and B, and with a separation between drifting
features corresponding to the period of A
\citep{mkl+04}. This provides clear evidence that pulsed radiation from
A is interacting with the magnetosphere
of B, supporting the picture that pulsar winds are
magnetically dominated ($\sigma \gg1$) in their inner regions,
as discussed in \S\ref{sec_young_pwn}.

Other binary systems also provide information on conditions
very close to the pulsar.  PSR~B1957+20 is in a circular 9.2-hour
orbit around a $\sim0.025$~M$_\odot$ companion, interaction with
which produces a termination shock just $1.5\times10^{11}$~cm
from the pulsar\footnote{On much larger scales, PSR~B1957+20 also
powers a bow shock, as shown in Figure~\ref{fig_w44_b1957}(b).}
\citep{pebk88}.  The X-ray flux from this nebula shows possible
modulation at the orbital period \citep{sgk+03}, as should result
from Doppler boosting of the flow around the companion \citep{at93}.
PSR~B1259--63 is in a highly eccentric 3.4-year orbit around a Be
star.  Near periastron, the pulsar is subject to a time-varying
external pressure, producing a transient X-ray/radio synchrotron
nebula plus TeV emission from IC scattering of light from the
companion star \citep[e.g.,][]{jbwm05,aaa+05c}.  At future periastra
of this system, coordinated X- and $\gamma$-ray observations
with {\em INTEGRAL}, {\em GLAST}\ and HESS can directly probe
particle acceleration in this pulsar's wind \citep{ta97,kbs99}.

%% file: summary.tex

\bigskip
\noindent
{\bf SUMMARY POINTS}

\addcontentsline{toc}{section}{SUMMARY POINTS}

\begin{enumerate}

\item A magnetized relativistic wind is the main reservoir for
a pulsar's rotational energy loss. The termination of this wind due to
surrounding pressure produces a pulsar wind nebula, usually observed as
a synchrotron nebula. A high velocity pulsar can also
produce a line-emitting optical bow shock where the pulsar wind
shocks surrounding gas.

\item Pulsar wind nebulae move through a series of distinct evolutionary
states, moderated by the pulsar's location (inside a SNR vs in
interstellar gas), the ambient conditions (cold ejecta vs shocked ejecta
vs ISM) and the Mach number of the pulsar (subsonic vs mildly supersonic
vs highly supersonic).

\item High resolution X-ray observations of young pulsars reveal
the imprint of the rotation axis on the morphology of
the surrounding PWN, in the form of equatorial tori, polar jets,
and overall elongation of the nebula. Using these structures, one can locate
the wind termination shock and can infer the composition, flow speed
and geometry of the pulsar's wind.

\item An increasing number of bow shocks are being found
around high-velocity pulsars.  These systems impose a second axis of
symmetry on the PWN, providing additional probes of the
pulsar's wind and environment.

\end{enumerate}